\documentclass[prd,aps,showpacs,nofootinbib]{revtex4}
\usepackage{amsmath}

\newcommand{\be}{\begin{equation}}
\newcommand{\ee}{\end{equation}}
\newcommand{\bq}{\begin{eqnarray}}
\newcommand{\eq}{\end{eqnarray}}

\begin{document}

\title{A Comment on the Topological Phase for Anti-Particles in a
Lorentz-violating Environment}
\author{H. Belich $^{a,e}$,T. Costa-Soares$^{c,d,e}$, M.M. Ferreira Jr.$%
^{b,e}$, J.A. Helay\"{e}l-Neto$^{c,d,e}$ and M. T. D. Orlando}
\affiliation{$^{a}${\small {Universidade Federal do Esp\'{\i}rito Santo (UFES),
Departamento de F\'{\i}sica e Qu\'{\i}mica, Av. Fernando Ferrari, S/N
Goiabeiras, Vit\'{o}ria - ES, 29060-900 - Brasil}}}
\affiliation{$^{b}${\small {Universidade Federal do Maranh\~{a}o (UFMA), Departamento de F%
\'{\i}sica, Campus Universit\'{a}rio do Bacanga, S\~{a}o Luiz - MA,
65085-580 - Brasil}}}
\affiliation{{\small {~}}$^{c}${\small {CBPF - Centro Brasileiro de Pesquisas F\'{\i}%
sicas, Rua Xavier Sigaud, 150, CEP 22290-180, Rio de Janeiro, RJ, Brasil}}}
\affiliation{$^{d}${\small {Universidade Federal de Juiz de Fora (UFJF), Col\'{e}gio T%
\'{e}cnico Universit\'{a}rio, av. Bernardo Mascarenhas, 1283, Bairro F\'{a}%
brica - Juiz de Fora - MG, 36080-001 - Brasil}}}
\affiliation{$^{e}${\small {Grupo de F\'{\i}sica Te\'{o}rica Jos\'{e} Leite Lopes, C.P.
91933, CEP 25685-970, Petr\'{o}polis, RJ, Brasil}}}
\email{belich@cce.ufes.br, tcsoares@cbpf.br, manojr@ufma.br,
helayel@cbpf.br, orlando@cce.ufes.br}

\begin{abstract}
Recently, a scheme to analyse topological phases in Quantum Mechanics by
means of the non-relativistic limit of fermions non-minimally coupled to a
Lorentz-breaking background has been proposed. In this letter, we show that
the fixed background, responsible for the Lorentz-symmetry violation, may
induce opposite Aharonov-Casher phases for a particle and its corresponding
antiparticle. We then argue that such a difference may be used to
investigate the asymmetry for particle/anti-particle as well as to propose
bounds on the associated Lorentz-symmetry violating parameters.
\end{abstract}

\maketitle

The Standard Model of Particle Physics is based on Lorentz- and $CPT$-
invariances as fundamental symmetries that have been confirmed in numerous
experiments \cite{barnet}-\cite{Clock}. Actually, the invariance under the
combined $CPT$- transformation is a consequence of first principles of
relativistic quantum field theory. The most immediate consequence of $\ CPT$
symmetry is the equality of mass and lifetime for a particle and its
corresponding anti-particle. The best tests in this direction come from the
limits on the mass difference between $K^{0}$ and $\overline{K^{0}}$ \cite%
{gibbons}, high precision measurements of the anomalous magnetic moments of
the electrons, positrons and mesons (confined in a Penning trap) \cite{dyck}%
, and clock-comparison experiments \cite{Clock}.

Lorentz-violating theories are presently studied as a possible extension of
the Standard Model of Particle Physics. This proposal has been pushed
forward by Colladay and Kosteletck\'{y} \cite{Kostelecky1}, who devised a
Standard Model Extension (SME)\ incorporating all tensor terms stemming from
the spontaneous symmetry breaking of a more fundamental theory. In this
case, an effective action breaks Lorentz symmetry at the particle frame, but
keeps unaffected the $SU(3)\times SU(2)\times U(1)$ gauge structure of the
underlying fundamental theory. This fact is of relevance in that it
indicates that the effective model may preserve some good properties of the
original theory, like causality, unitarity and stability. \

In the context of gauge theories endowed with Lorentz violation, some
efforts have been recently devoted\textbf{\ }to investigate interesting
features of relativistic quantum-mechanical models{\large \ }involving\ the
presence of fermions. Indeed, considering\textbf{\ }the Dirac equation
enriched by of a sort of non-minimal coupling, significant consequences on
the particle behavior has been observed, as pointed out in the works of ref.
\cite{acnaominimo}. In these papers, the analysis of the non-relativistic
regime of the Dirac equation has revealed that topological quantum phases
are induced\ whenever the fermion field is coupled to the fixed background
and the gauge field in a non-minimal way. More specifically\textbf{,} it has
been found out that a neutral particle acquires a magnetic moment (induced
by the background), which originates the Aharonov-Casher (AC) phase subject
to the action of an external electric field \cite{casher}. It is worth
stressing that the standard Aharonov-Casher phase is interpreted as a
Lorentz change in the observer frame. In our proposal, it rather emerges as
a phase whose origin is ascribed to the presence of a privileged direction
in the space-time, set up by the fixed background. Since in this kind of
model Lorentz invariance in the particle frame is broken, the
Aharonov-Casher effect could not here be obtained by a suitable Lorentz
change\ in the observer frame.

In this letter, rather than focusing the quest for $CPT$- violation on
possible particle/anti-particle mass or lifetime differences, we elect the
Aharonov-Casher effect as an effective probe to detect
particle/anti-particle asymmetry in a Lorentz violating environment. Based
on the result of ref\textbf{.}{\large \ }\cite{acnaominimo}\textbf{,} we
show that particles and anti-particles, even if spinless and neutral, may
pick up opposite AC-phases in their wave functions. This may lead to
interference \ effects that could be used to verify potential
particle/anti-particle asymmetries as a consequence of Lorentz and/or $CPT$
violation induced by some background. This outcome may be also used to set
tight bounds on the $CPT$-violating parameters. Our letter is outlined as
follows. First, we begin by reviewing the standard minimal coupling of a
charged anti-particle, which gives the Aharonov-Bohm (AB) phase. We then
observe that for the non-relativistic (Pauli) Hamiltonian, the AB phase for
a particle and its anti-particle presents no modification. Afterwards, we
contemplate the standard non-minimal coupling (the Pauli interaction) that
gives rise to the Aharonov-Casher phase for a non-relativistic neutral
particle endowed with magnetic dipole moment, showing that in this case the
AC phase does not undergo any change under $CPT$ transformation as well.
Finally, we investigate the case of main interest:\ some possible
Lorentz-violating non-minimal couplings that induce AC phases. In the
situations considered here, it is demonstrated that the AC phases for a
particle and its anti-particle exhibit opposite signs.

Our starting point is an investigation on the Aharonov-Bohm phase induced
for the anti-particle by the usual minimal coupling to the electromagnetic
field, which respects the\textbf{\ }$CPT$\textbf{-\ }symmetry. We then write
the gauge invariant Dirac equation for the anti-particle, which may be
obtained from the standard Dirac equation after a $CPT$ transformation. The
spinor wave function transforms under $C$, $P$ and $T$ \ operations as
follows: $\Psi \overset{\mathcal{P}}{\longrightarrow }\eta _{P}\gamma
^{0}\Psi $, $\Psi \overset{\mathcal{C}}{\longrightarrow }\eta _{C}C\overline{%
\Psi }^{t}$ and $\Psi \overset{\mathcal{T}}{\longrightarrow }\eta _{T}\gamma
^{1}\gamma ^{3}\Psi ^{\ast }$, where the $\eta $'s correspond to the phases
associated with each transformation and $C$ is the charge conjugation
matrix, given in this representation\footnote{%
Here, we adopt the usual Dirac representation for the $\gamma $-matrices.}
by $C=i\gamma ^{0}\gamma ^{2}$. Considering the overall effect of these
transformations\textbf{,} the spinor function describing the anti-particle
changes as below:
\begin{equation}
\Psi _{CPT}=\gamma _{5}\Psi .
\end{equation}%
Moreover, under $CPT-$transformation\footnote{%
We adopt the prescriptions found in ref. \cite{bjorken}.} it holds: $%
\partial _{\mu }\overset{CPT}{\rightarrow }-\partial _{\mu }$, \ $A_{\mu }%
\overset{CPT}{\rightarrow }-A_{\mu },$\ $F_{\mu \nu }\overset{CPT}{%
\rightarrow }F_{\mu \nu }.$ So, for the anti-particle, the Dirac equation
reads as:
\begin{equation}
\left( i\gamma ^{\mu }\partial _{\mu }+m-e\gamma ^{\mu }A_{\mu }\right)
\gamma _{5}\Psi =0.
\end{equation}%
The first case to be discussed here\ concerns the AB phase; we take both
particle and anti-particle are minimally coupled to the electromagnetic
vector potential via the covariant derivative with minimal coupling. To work
out the non-relativistic limit of the Dirac equation for the anti-particle,
the spinor $\Psi $\ should be written in terms of the so-called small and
large components, as it is usually done. Thereby, there appear two coupled
equations for the 2-component spinors, that, once decoupled, yield{\large \ }%
the non-relativistic Hamiltonian,
\begin{equation}
H=-\frac{1}{2m}\overrightarrow{\Pi }^{2}-e\varphi +\frac{e}{2m}%
\overrightarrow{\sigma }\cdot (\overrightarrow{\nabla }\times
\overrightarrow{A}),
\end{equation}%
and the canonical conjugated momentum for the strong component, $%
\overrightarrow{\Pi }=(\overrightarrow{p}-e\overrightarrow{A}).$ This puts
in evidence that the energy of the anti-particle (before reinterpretation)
has opposite sign in comparison with the energy for the corresponding
particle, as a consequence of $CPT$ transformation. However, the phase
factor associated with the Aharonov-Bohm effect does not change its sign.
Indeed, the bilinear term\ ($A^{\mu }J_{\mu })$\ of the Lagrangian, where
the AB phase stems from, remains invariant.

The Aharonov-Casher phase for the anti-particle can be analysed on the same
grounds. To work it out, we consider an electrically neutral particle (not
necessarily a Majorana fermion), described by a spinor, $\Psi $, whose
behavior is governed by the Dirac equation non-minimally coupled to gauge
field:
\begin{equation}
(i\gamma ^{\mu }\partial _{\mu }-m+\frac{1}{2}\mu F^{\mu \nu }\Sigma _{\mu
\nu })\Psi =0,
\end{equation}%
where the non-minimal term under $CPT$ transformation\ goes as: $\mu F^{\mu
\nu }\Sigma _{\mu \nu }\overset{CPT}{\rightarrow }\mu F^{\mu \nu }\Sigma
_{\mu \nu }.$ To isolate the AC-phase for the anti-particle, the magnetic
field is set to zero\textbf{,} $F^{ij}=0$. After simple algebraic
manipulations, $F^{\mu \nu }\Sigma _{\mu \nu }$ turns into $i\mu E^{i}\gamma
_{i}\gamma _{0}$, so that the Dirac equation takes on the form: $\left(
-i\gamma ^{\mu }\partial _{\mu }-m-i\mu E^{i}\gamma ^{i}\gamma ^{0}\right)
\gamma _{5}\Psi =0.$ Again, to compute the non-relativistic limit, the
spinor $\Psi $\ is split into strong and weak components, yielding the
following Hamiltonian:
\begin{equation}
H=\frac{1}{2m}\left( \overrightarrow{p}-\overrightarrow{E}\times
\overrightarrow{\mu }\right) ^{2}-\frac{\mu ^{2}\overrightarrow{E}^{2}}{2m},
\end{equation}%
while\textbf{\ }$\overrightarrow{\Pi }=(\overrightarrow{p}-\overrightarrow{E}%
\times \overrightarrow{\mu })$\textbf{\ }is the conjugate momentum. This
shows that, as in the case of the AB phase, the phase acquired by the
anti-particle is not modified. This phase invariance is obviously a
consequence of\textbf{\ }the Lorentz and $CPT$ symmetries of the non-minimal
coupling above.

We now go on to investigate the effect of a unusual non-minimal coupling on
the topological A-Chaser phase\textbf{.}{\large \ }As it has been recently
shown \cite{acnaominimo}, the Lorentz-breaking background\textbf{\ (}$v^{\mu
})$\textbf{\ }may induce the appearance of an AC phase for neutral fermionic
particles, whenever it is non-minimally coupled to the gauge and spinor
fields by means of the covariant derivative\textbf{, }$D_{\mu }=\partial
_{\mu }+$\textbf{\ }$ieA_{\mu }+igv^{\nu }F_{\mu \nu }^{\ast },$\ properly
inserted into the Dirac equation. Here\textbf{, }$F_{\mu \nu }^{\ast }$%
\textbf{\ }denotes the dual tensor.\ The non-minimal coupling introduced
here in the covariant derivative, $D_{\mu }$, has actually been contemplated
in the works of ref. \cite{blum2}, where the authors propose an exhaustive
list of terms that may respond for the breaking of Lorentz symmetry. Indeed,
in the equation (1) of \cite{blum2}, if we identify $a_{\mu }^{c}\equiv
igv^{\nu }F_{\mu \nu }^{\ast }$, we recover the non-minimal coupling to $%
v^{\mu }$ that appears in our covariant derivative.

Another motivation to introduce a term like $igv^{\nu }F_{\mu \nu }^{\ast }$
in the covariant derivative is that, in the case $v^{\mu }$ is space-like,
this breaking term may naturally induce the non-minimal coupling considered
by Khare and Paul \cite{paul}, where the term in $F_{\mu }^{\ast }$ $\left(
\mu =0,1,2\right) $ is responsible for the magnetic moment of scalar
particles in $\left( 1+2\right) $ dimensions.

The conjugate momentum obtained is then given as\textbf{:} \ $%
\overrightarrow{\Pi }=(\overrightarrow{p}-e\overrightarrow{A}+gv_{0}%
\overrightarrow{B}-g\overrightarrow{v}\times \overrightarrow{E})$. The
presence of the term $g\overrightarrow{v}\times \overrightarrow{E}$ \ leads
to the\textbf{\ }Aharonov-Casher effect, with the background 3-vector $%
\overrightarrow{v}$ playing the role of a sort of magnetic dipole moment, $%
\left( \overrightarrow{\mu }=g\overrightarrow{v}\right) $, that generates
the AC phase associated with the wave function of a neutral test particle ($%
e=0$) \cite{acnaominimo}. For the case of a charged particle,\ we should
remark that there appear both AB and AC phases simultaneously. The presence
of the background vector $v^{\mu }$ endows the space-time with the feature
that, wherever the particle is placed (and it has the property of coupling
to $v^{\mu }$ with a coupling constant $g$ ), it acquires an intrinsic
vector attribute given by $g\overrightarrow{v}$; since it couples to an
electric field to give an AC phase factor, we are led to interpret the
quantity $g\overrightarrow{v}$ as being a sort of magnetic moment. It is
true that \ it is fixed by the background, and it is not exactly like the
magnetic moment associated to the spin.

Considering now the Dirac equation supplemented by the non-minimal coupling
above, $\left( i\gamma ^{\mu }\partial _{\mu }-m-e\gamma ^{\mu }A_{\mu
}-g\gamma ^{\mu }v^{\nu }F_{\mu \nu }^{\ast }\right) \Psi =0$, and adopting
the standard viewpoint that the background vector, $v^{\mu }$, is not
transformed, we can show, by performing the $C,$\ $P$\ and $T$\
transformations, that the anti-particle, described by the wave function $%
\Psi _{CPT}=\gamma _{5}\Psi ,$\ satisfies the eq.%
\begin{equation}
\left( i\gamma ^{\mu }\partial _{\mu }+m-e\gamma ^{\mu }A_{\mu }+g\gamma
^{\mu }v^{\nu }F_{\mu \nu }^{\ast }\right) \gamma _{5}\Psi =0,
\label{Dirac1B}
\end{equation}%
This clearly shows the change in the sign of the interaction term that
yields the AC phase. We shall now focus on this point.

Considering \ the non-relativistic limit of the Dirac equation given above (%
\ref{Dirac1B}), it results that the strong spinor component satisfies the
following non-relativistic Hamiltonian:

\begin{equation}
H=-\frac{1}{2m}\overrightarrow{\Pi }^{2}-e\varphi -\frac{e}{2m}%
\overrightarrow{\sigma }\cdot (\overrightarrow{\nabla }\times
\overrightarrow{A})-\frac{1}{2m}gv^{0}\overrightarrow{\sigma }\cdot (%
\overrightarrow{\nabla }\times \overrightarrow{B})-\frac{g}{2m}%
\overrightarrow{\sigma }\cdot \overrightarrow{\nabla }\times (%
\overrightarrow{v}\times \overrightarrow{E}),
\end{equation}%
where the canonical generalized momentum is given by\textbf{: }$%
\overrightarrow{\Pi }=(\overrightarrow{p}-e\overrightarrow{A}-gv^{0}%
\overrightarrow{B}+g\overrightarrow{v}\times \overrightarrow{E}).$\textbf{\ }%
Such a momentum is to be compared with the momentum associated with the
particle, given previously, which exhibits two terms with opposite signs$.$
This clearly shows that the particle and its corresponding anti-particle
acquire opposite AC phases as due to the non-minimal coupling we are
considering.

To stress this result, we present below another case where Lorentz
invariance is violated by an external tensor, $T_{\mu \nu }$, also
non-minimally coupled, but not through a covariant derivative. The
motivation to introduce the tensor $T^{\mu \nu }$ with the non-minimal
coupling given by $\lambda _{1}$ has already been discussed by Kosteleck\'{y}
in the work of ref. \cite{kosta}. We however add another non-minimal term
that couples $T^{\mu \nu }$ also to $F^{\mu \nu }$ with coupling parameter $%
\lambda _{2}$. This term has not been discussed in \cite{kosta} and we have
included it because it allows us to understand how the background electric
and magnetic fields may influence on the dynamics of the fermion by means of
their coupling to the background tensor $T^{\mu \nu }$. According to the
results of the works of ref. \cite{Kostelecky1}, both these terms are $CPT$
even.

In this situation, the Dirac equation is given by
\begin{equation}
(i\gamma ^{\mu }D_{\mu }-m+i\lambda _{1}T_{\mu \nu }\Sigma ^{\mu \nu
}+i\lambda _{2}T_{\mu \kappa }F^{\kappa }\text{ }_{\nu }\Sigma ^{\mu \nu
})\Psi =0,  \label{tensor11}
\end{equation}%
where $\Sigma ^{\mu \nu }=i[\gamma ^{\mu },\gamma ^{\nu }]/2,$ $\lambda _{1}$
and $\lambda _{2}$ are the coupling constants.\textbf{\ }The analysis of the
non-relativistic regime shows that the AC phase for the particle is related
to the generalized momentum\textbf{\ }$\overrightarrow{\Pi }=(%
\overrightarrow{p}-e\overrightarrow{A}-4\lambda _{1}\overrightarrow{T}%
_{1}-2\lambda _{2}\overrightarrow{T}_{2}\times \overrightarrow{E})$\textbf{,
}with\textbf{\ }$\overrightarrow{T}_{1}=T_{0i},\overrightarrow{T}_{2}=T_{ij}.
$\ By suitably performing the $C$, $P$ and $T$ transformations on the spinor
wave function and the electromagnetic field in eq. (\ref{tensor11}), without
however changing $T_{\mu \nu }$(the background is taken to be invariant
under C,P and T), we get to the conclusion that the anti-particle wave
function is a solution to the equation%
\begin{equation}
(i\gamma ^{\mu }D_{\mu }+m-i\lambda _{1}T_{\mu \nu }\Sigma ^{\mu \nu
}-i\lambda _{2}T_{\mu \kappa }F^{\kappa }\text{ }_{\nu }\Sigma ^{\mu \nu
})\gamma _{5}\Psi =0.
\end{equation}%
This ensures therefore that the anti-particle will develop an opposite AC
phase. Though, the terms brought about $T_{\mu \nu }$ are $CPT$ -even. We
understand that, due to the fact that they break Lorentz symmetry, they
induce an opposite AC phase for the anti-particle

These results suggest\textbf{\ }that Lorentz-symmetry and $CPT$ violations
may be probed by experiments based on the phase interference of particle and
anti-particle wave functions in the context of a typical Aharonov-Casher
set-up. Indeed, in a non-simply connected region of space, where a beam of
neutral particles is subject to a radial electric field, a shift of fringes
is observed after the electric field is turned on (Aharonov-Casher effect).
In the case this same experiment is performed with an anti-particle beam,
the variation of the interference pattern is expected to show up in\ the
opposite sense. The observation of such an effect might then be used to
dictate bounds on the magnitude of the tensor terms\textbf{\ (}$v^{\mu
},T_{\mu \nu })$ responsible for the violation of the Lorentz symmetry%
\textbf{. }This question demands a special investigation and we shall be
concerned with it in a forthcoming work, focused on the task of determining
bounds on the objects $v^{\mu }$ and $T^{\mu \nu }$ responsible for the
breaking of Lorentz symmetry \cite{nos2}. We do not know of any experimental
set-up that infers about AC phases for anti-particles. The question may also
be studied for vector bosons and other particles with spin higher than $%
\frac{1}{2}$. This investigation is under consideration and we shall report
on it elsewhere \cite{Novo}.

\section{Acknowledgements}

MMFJr, JAH-N and MTDO are grateful to CNPq for the invaluable financial help.

\end{document}